\documentclass[%
 reprint,longbibliography,preprintnumbers,
nofootinbib,
 amsmath,amssymb,
 aps,
prb,
]{revtex4-2}
\pdfoutput=1
\usepackage[utf8]{inputenc}
\usepackage{flushend}
\usepackage{dcolumn}
\usepackage{bm}
\usepackage{balance}
\usepackage{amssymb}

\usepackage[normalem]{ulem}

\usepackage[colorlinks = true,
            linkcolor = blue,
            urlcolor  = blue,
            citecolor =green,
            anchorcolor = blue]{hyperref}
\usepackage{verbatim}
\usepackage{color,ulem}
\usepackage[english]{babel}

\usepackage[utf8]{inputenc}
\input Starburst.fd
\newcommand*\initfamily{\usefont{U}{Starburst}{xl}{n}}\initfamily

\newcommand{\beq}{\begin{eqnarray}}
\newcommand{\eeq}{\end{eqnarray}}
\usepackage{amsmath}
\usepackage{tikz}
\usetikzlibrary{decorations.pathmorphing}
\usetikzlibrary{shapes.misc}
\tikzset{cross/.style={cross out, draw=black, minimum size=8*(#1-\pgflinewidth), inner sep=0pt, outer sep=0pt},
cross/.default={1pt}}
\usetikzlibrary{patterns,math}
\begin{document}

\title{Theory of superconductivity in thin films under an external electric field}

\author{Alessio Zaccone$^{1,2}$}
\author{Vladimir M. Fomin$^{3,4}$}

\affiliation{$^1$  Department of Physics ``A. Pontremoli'', University of Milan, via Celoria 16,
20133 Milan, Italy}
\affiliation{$^2$ 
Institut für Theoretische Physik, University of G{\"o}ttingen,
Friedrich-Hund-Platz 1
37077 G{\"o}ttingen, Germany}
\affiliation{$^3$ 
Institute for Emerging Electronic Technologies
Leibniz IFW Dresden
Helmholtzstrasse 20, D-01069 Dresden, Germany}
\affiliation{$^4$ 
Laboratory of Physics and Engineering of Nanomaterials
Department of Theoretical Physics
Moldova State University
str. Alexei Mateevici 60, MD-2009 Chişin\u{a}u, Republic of Moldova}

 \vspace{1cm}

\begin{abstract}
The supercurrent field effect is experimentally realized in various nano-scale devices, based on the superconductivity suppression by external electric fields being effective for confined systems. In spite of intense  research, a microscopic theory of this effect is missing.
Here, a microscopic theory of phonon-mediated superconductivity in thin films under an external electric field is presented, which accounts for the effect of quantum confinement on the electronic density of states, on the Fermi energy, and on the electron Coulomb repulsion. By accounting for the complex interplay between quantum confinement, the external static electric field, the Thomas-Fermi screening in the electron-phonon matrix element, and the effect of confinement on the Coulomb repulsion parameter, the theory predicts the critical value of the external electric field as a function of the film thickness, above which superconductivity is suppressed. In particular, this critical value of the electric field is exponentially lower the thinner the film, in agreement with recent experimental observations. Crucially, this effect is predicted by the theory when both Thomas-Fermi screening and the Coulomb pseudopotential are taken into account, along with the respective dependence on the thin film thickness. This microscopic theory opens up new possibilities for the supercurrent field effect and for electric-field gated quantum materials.
\end{abstract}

\maketitle
\section{Introduction}
The ability to modulate an electric current via an externally-imposed electric field is at the heart of field effect transistors \cite{Bardeen_transistor} and of all electronic industry. 
The same effect cannot be reproduced in metallic materials, due to the screening of the external electric field (EF) by the mobile charges \cite{Lang_Kohn_1,Jackson}.

The effect of the EF-driven suppression of superconductivity was unveiled in pioneering works by Giazotto and his collaborators in titanium  nanostructures: nanowires \cite{Giazotto2018}, Dayem bridges \cite{Paolucci,Paolucci2019,Rocci2020} and superconducting quantum interference devices (SQUIDs) \cite{Paolucci2019-2}. The EF-induced quenching of the superconducting state was observed in different superconducting materials, mostly purely metallic, for example: proximitized Al/Cu/Al nanojunctions \cite{DeSimoni2019}, aluminium nanodevices with a single back-gate \cite{Giazotto3} or side-gate {\cite{Bours}}, vanadium Dayem nano-bridges \cite{Puglia2020}, niobium gate-controlled transistors \cite{DeSimoni2020}, and WC nanowires \cite{Orus2021}. 

A reasonable phenomenological description of the field effect on supercurrent in Ti \cite{Giazotto2018} and WC \cite{Orus2021} superconductor nanowires within the Landau-Ginzburg theory has been achieved based on the assumption of the suppression of the superconducting order parameter near the edges of the nanowire. 

The microscopic interpretation of the EF effect on superconductivity in nanostructures has been the subject of a vivid debate on several hypotheses of intrinsic and extrinsic effects for various confinement geometries and in different superconducting materials. A few of them are listed below. The transition from superconducting to normal state was ascribed to injection of high-energy electrons from the gate electrodes to the TiN nanowire triggering generation of quasiparticles which, at sufficiently large currents,  lead to heating \cite{Ritter}. The similar mechanism was assumed to explain the electron tunneling spectroscopy experiments in titanium nanowires \cite{Alegria} and in a vanadium waveguide resonator \cite{Golokolenov}. A theoretical approach for thin superconducting films was based on EF-tuned spin-orbit polarization at the surface, giving rise to modulation of the phase and amplitude of the superconducting order parameter \cite{Mercado}. A similar theoretical model was presented for thin crystalline superconductors, in which the EF-impact was attributed to a local modification of the density of states of the material through Rashba-spin-oirbit-interaction-like surface effects \cite{Chirolli}. More recently, a Sauter--Schwinger effect in BCS superconductors was proposed: EF can generate two coherent excitations from the superconducting ground-state condensate, which form a new, macroscopically coherent and dissipationless, state that gives rise to weakening of the superconducting state \cite{Solinas}. However, a microscopic theory of superconductivity suppression in nanostructures that simultaneously takes into consideration the effects of quantum confinement in interplay with the external EF, is still an open issue.

Here, we develop such a microscopic theory within the Barden--Cooper--Schrieffer (BCS) framework, which takes into account the effect of thin-film confinement on the Fermi energy and density of states at Fermi level using recent theoretical developments \cite{Travaglino}, and successfully predicts that decreasing values of critical EF $E_{cr}$ are required to suppress superconductivity of thin films at decreasing values of the film thickness $L$.
We work under the usual assumptions of the Migdal-Eliashberg theory and under the assumption that changes in carrier density due to a shift of the Fermi level can be neglected. This implies that the current theory cannot be applied to Bose-Einstein condensates (BEC) or semi-metals, which are left for a future work.

\section{Microscopic theory}
\subsection{Eliashberg formalism}
Within the Eliashberg theory of electron-phonon superconductivity, the so-called Eliashberg function represents the spectrum of phonon energies which are responsible for electron pairing:
\begin{equation}
    \alpha^2F(\boldsymbol{k},\boldsymbol{k'},\omega)\equiv N(\epsilon_F)|g_{\boldsymbol{k}\boldsymbol{k'}}|^2 \mathcal{B}(\boldsymbol{k}-\boldsymbol{k'},\omega), \label{start point}
\end{equation}
where $N(\epsilon_F)$ is the electronic density of states (DOS) at the Fermi energy $\epsilon_F=\mu$, where $\mu$ is the chemical potential, $g_{\boldsymbol{k},\boldsymbol{k'}}$ is the electron-phonon matrix element and $k,k'$ are the wave-vectors of the paired electrons.
The phonon propagator has the typical Lorentzian form for the spectral density
\begin{equation}
    \mathcal{B}(q,\omega)=-\frac{1}{\pi} \mathrm{Im}\mathcal{G}(\omega,q)=\frac{\omega\Gamma(q)}{\pi[(\omega^2-\Omega^2(q))^2+\omega^2\Gamma^2(q)]}.\label{spec}
\end{equation}
where $q$ indicates the phonon momentum.

Upon summing over all electronic momentum states, one obtains \cite{Carbotte2003}:
\begin{equation}
\alpha^{2}F(\omega)=\frac{1}{N(\epsilon_F)^2}\sum_{\boldsymbol{k},\boldsymbol{k'}}\alpha^{2}F(\boldsymbol{k},\boldsymbol{k'})\delta(\epsilon_k - \epsilon_F)\delta(\epsilon_{k'} - \epsilon_F)
\label{double-sum}
\end{equation}
where we recall that $\sum_{k}\delta(\epsilon_k - \epsilon_F)\propto N(\epsilon_F)$.
Upon replacing Eq. \eqref{start point}, we then have:
\begin{equation}
\alpha^{2}F(\omega)=\frac{|g_{\boldsymbol{k}\boldsymbol{k'}}|^2}{N(\epsilon_F)}\sum_{\boldsymbol{k},\boldsymbol{k'}} \mathcal{B}(\boldsymbol{k}-\boldsymbol{k'},\omega)\delta(\boldsymbol{k} - \epsilon_F)\delta(\boldsymbol{k'} - \epsilon_F).
\label{double-sum 2}
\end{equation}
The sum in momentum space in Eq. \eqref{double-sum} can be written
\begin{equation}  \frac{1}{N}\sum_{\boldsymbol{k}}...\rightarrow \int ...N(\epsilon) d\epsilon
\label{sum}
\end{equation}

The Eliashberg function is then used as input to compute the electron-phonon coupling constant $\lambda$ \cite{Carbotte2003}:
\begin{equation}
\lambda = 2 \int_{0}^{\infty}\frac{\alpha^{2}F(\omega)}{\omega}d\omega. \label{lambda}
\end{equation}
By further assuming a constant electronic density of states approximation, extended over an infinite band, and a single optical phonon as the mediator with frequency $\omega_E$, i.e. $\mathcal{B}(\boldsymbol{k}-\boldsymbol{k'},\omega) \equiv \delta(\omega - \omega_E)$ \cite{Carbotte2003}, one obtains \cite{Carbotte2003}
\begin{equation}
\lambda = 2 N(\epsilon_F)|g_{\boldsymbol{k}\boldsymbol{k'}}|^2 / \omega_E. \label{lambda_final}
\end{equation}

Next we need to evaluate the electron-phonon matrix element $g_{\boldsymbol{k}\boldsymbol{k'}}$ and in particular its dependence on $\epsilon_F$ and $N(\epsilon_F)$.

The standard Bardeen-Pines formula for the electron-phonon matrix element reads (in their notation) as \cite{Bardeen_Pines,Kittel_Theory}:
\begin{equation}
v_{q}^{Z}=\frac{4\pi Ze^2}{q}\left(\frac{N}{M}\right)^{1/2}
\end{equation}
where $q=|\boldsymbol{k}-\boldsymbol{k}'|$, $Z$ is the atomic number, $N$ the ion density, $M$ the ionic mass, $e$ the electron charge.

\subsection{External electric field and Thomas-Fermi screening}
In the presence of an external DC electric field (EF) applied transversely across the thin film, two main effects  are brought about by the EF:\\
(i) the Fermi sphere, as a whole, is shifted, in $k$-space, in the direction of the applied EF by a quantity $m v_{d}/\hbar$, where $v_{d}$ is the magnitude of the drift velocity of the conduction electrons in the applied EF and $m$ is the electron mass. This effect has no influence at all on the quantities relevant for superconductivity since the Fermi energy and the structure and topology of electron states in $k$-space are not altered;\\
(ii) screening effects become important \cite{Kittel}, and need to be properly taken into account within a microscopic theory of superconductivity.

Based on these considerations, we shall then model the electron-phonon interaction accordingly, and work in the regime of non-negligible Thomas-Fermi screening as appropriate for a metallic material subjected to an external static EF.

According to Engelsberg and Schrieffer  \cite{PhysRev.131.993}, screening can be taken into account by considering the dielectric constant of a free electron gas immersed in a background positive charge due to the ions, as follows:
\begin{equation}   g_{\boldsymbol{k}\boldsymbol{k'}}^2=|v_{q}^{Z}/\epsilon(0,q)|^2=\left|\frac{4\pi Ze^2}{q}\left(\frac{N}{M}\right)^{1/2}\frac{q^2}{q^2+k_{s}^{2}}\right|^2, \label{gq}
\end{equation}
where $\epsilon(0,q)=1+\frac{k_{s}^{2}}{q^{2}}$ is the standard static dielectric constant computed by solving the electrostatic Poisson-equation problem  for an electron gas in presence of the positive background charge of the lattice \cite{Kittel}, and $k_s$ denotes the screening wave-vector.

For a quantum Fermi gas at low temperature, the screening wave-vector is given by the Thomas-Fermi wave-vector as \cite{Ashcroft}:
\begin{align}
k_{TF}^{2}=4 \left(\frac{3 n}{\pi}\right)^{1/3}
\label{Thomas-Fermi}
\end{align}
which is the inverse Thomas-Fermi screening length, with $n$ the number density of electrons. For a non-degenerate electron gas (the Debye-H{\"u}ckel screening), $k_{s}=\sqrt{\frac{4 \pi  e^2 n}{k_{B}T}}$, with $e$ the electron charge and $k_{B}$ the Boltzmann constant.

When the screening wavevector is large compared to the momentum of the mediating phonon, $k_s \equiv k_{TF} \gg q$, one can approximate the above expression Eq. \eqref{gq}, as simply:
\begin{equation}
g_{\boldsymbol{k}\boldsymbol{k'}}=\frac{c}{k_{TF}^{2}}|q|.
\label{e-ph_simple0}
\end{equation}
At thermodynamic equilibrium, the electron concentration is related to the chemical potential, at $T=0$, via the relation \cite{Ashcroft}
\begin{equation}
n \sim \mu^{3/2}
\end{equation}
Upon substituting into Eq. \eqref{Thomas-Fermi}, one gets:
\begin{equation}
k_{TF}^{2} \sim n^{1/3} \sim \mu^{1/2}=A \, \epsilon_{F}^{1/2}
\end{equation}
where $A$ is a constant. 
We now substitute this into Eq. \eqref{e-ph_simple0}, and obtain:
\begin{equation}
g_{\boldsymbol{k}\boldsymbol{k'}}^2=\frac{c}{A}\frac{|q|}{\epsilon_{F}^{1/2}}=\frac{c'}{\epsilon_{F}^{1/2}}|q|.
\label{e-ph_simple}
\end{equation}

According to Refs. \cite{PhysRev.131.993,nature06874,PhysRevLett.115.176401}, the momentum-dependence of the electron-phonon matrix elements for optical phonons can be safely neglected in good approximation. 

We can thus replace this result in Eq. \eqref{lambda_final}, and, in the case of strong screening, obtain:
\begin{equation}
\lambda = \frac{2\,D N(\epsilon_F)}{\epsilon_{F}^{1/2}}.
\label{lambda screening}
\end{equation}
where $D$ is a constant independent of $\epsilon_F$, whereas for systems where the Thomas-Fermi wave vector is small compared to the momentum of the pairing phonon, $k_{TF}\ll q$, we have:
\begin{equation}
\lambda = 2\,D' N(\epsilon_F).
\label{lambda no screening}
\end{equation}

Whether the screening wave vector is large compared to the wavevector of the phonon mediating the Cooper pair is of course something that depends on the microscopic details of the pairing in a given system. In any case, we consider both situations in the following.

\subsection{Critical superconducting temperature}
Finally, to evaluate the critical superconducting temperature $T_c$, we use the Allen-Dynes formula \cite{PhysRevB.12.905}:
\begin{equation}
      T_c\,=\,\frac{f_1\,f_2\,\omega_{log}}{1.2}\,\exp\left(-\frac{1.04\,(1+\lambda)}{\lambda-u^\star\,-\,0.62\,\lambda\,u^\star}\right)\label{allenformula}
\end{equation}
where 
\begin{equation}
\omega_{log}=\exp \left(\frac{2}{\lambda}\int_{0}^{\infty} d\omega \frac{\alpha^{2}F(\omega)}{\omega} \ln \omega \right)
\end{equation}
represents the characteristic energy scale of phonons for pairing in the strong-coupling limit, $f_1$, $f_2$ are semi-empirical correction factors of order unity, as defined in ~\cite{PhysRevB.12.905}, and $u^\star$ is the Coulomb repulsion parameter.
As a consistency check, the dependence of $\lambda$ on $N(\epsilon_F)$ in Eq. \eqref{lambda} is such that, within Eq. \eqref{allenformula}, one correctly recovers the dependence of $T_c$ on $N(\epsilon_F)$ as predicted by the BCS formula in the weak-coupling limit.

\subsection{Confinement effects of thin film thickness}
We should now implement the dependence on the film thickness $L$, and remark that the film thickness $L$ is much larger than the skin depth $\delta$, in most situations of interest.
In spite of this, as we shall see shortly below, it is the value of the film thickness $L$ that strongly affects the density of states at Fermi level, $N(\epsilon_F)$, due to the effects of quantum confinement in the thin film geometry. 
As demonstrated in Ref. \cite{Travaglino}, the geometry and topology of occupied electronic states in $k$-space changes dramatically upon decreasing the film thickness $L$. This, in turn, leads to a change in the electronic DOS from the standard square-root of energy to a linear-in-energy law once $L$ decreases below a critical thickness value $L_c = \sqrt[3]{\frac{2\pi}{n}}$. The mathematical predictions were quantitatively verified for experimental data of crystalline thin films and were able to quantitatively reproduce the trend of $T_c$ vs $L$, including the maximum of $T_c$ at $L=L_c$.
In a nutshell, two symmetrical (with respect to the center of the Fermi sphere) spherical cavities of forbidden states (due to thin-film confinement) are predicted to grow inside the Fermi sphere upon decreasing $L$.
The two spheres of forbidden states grow further up to the point, at $L=L_c$, where the spherical Fermi surface is ``punched'' and a topological transition occurs from the trivial sphere $\pi_{1}(S^{2})=0$ to a non-trivial Fermi surface with homotopy group 
$\pi_{1} \simeq \pi_{1}(S^{1})=\mathbb{Z}$.
In the regime $L>L_c$ as the two cavities grow with decreasing $L$, a redistribution of states density from the interior towards the Fermi surface occurs, which increases the DOS at the Fermi level. For $L<L_c$, instead, as $L$ decreases further, the Fermi surface of the system grows and the states become more spread out on the Fermi surface, hence the DOS at Fermi level here decreases upon further decreasing $L$. For Pb thin films, in Ref. \cite{Travaglino} it was found that $L_c \approx 5${\AA}.

Analytically, from the theory one obtains \cite{Travaglino}:
\begin{align}
      N(\epsilon_F) &= N^{bulk}(\epsilon_{F}) \left(1 + \frac{2}{3} \frac{\pi}{n L^3}\right)^{1/3}, ~~L>L_c  \label{Fermi_L_large0}\\
      N(\epsilon_F)&= 2\frac{V m \sqrt{L n}}{\sqrt{2} \pi^{3/2}\hbar^2},~~~~~L<L_c.
 \label{Fermi_L_large}
  \end{align}
Furthermore, also the Fermi energy changes as a function of the film thickness $L$, in the two regimes, according to the following expressions:
\begin{align}
    \epsilon_F &= \epsilon_F ^{bulk} \left(1+\frac{2}{3} \frac{\pi}{n L^3}\right)^{2/3}, ~~L>L_c \label{Fermi_L_small0}\\ 
   \epsilon_F &=  \frac{\hbar^2}{m}\left[\frac{(2\pi)^3\,n}{ L}\right]^{1/2},~~~~~L<L_c.
   \label{Fermi_L_small}
\end{align}

Since $L_c \approx 4-5${\AA} \cite{Travaglino} is of the same order of magnitude as the skin depth $\delta$, we shall start by focusing on the regime $L>L_c$.

\section{Results: thickness-dependent suppression of superconductivity}
\subsection{Derivation with thickness-independent Coulomb repulsion}
According to the Allen-Dynes formula Eq. \eqref{allenformula}, the superconductivity would be suppressed when the denominator of the argument of the exponential goes to zero, i.e. for a value of electron-phonon coupling constant given by:
\begin{equation}
\lambda_{cr} = \frac{u^\star}{1-0.62 u^\star }.
\end{equation}
As argued above, the Coulomb parameter $u^\star$ is not expected to vary with confinement $L$. Hence, possible onset/suppression of superconductivity may occur at a critical value of film thickness $L_{cr}$, which corresponds to $\lambda_{cr}$. The existence, or not, of this $L_{cr}$ can be verified as follows.
Inserting our result for $\lambda$ in the case where screening is large compared to the pairing wave vector, Eq. \eqref{lambda}, we thus obtain the condition:
\begin{equation}
 \frac{2\,D\, N(\epsilon_F)}{\epsilon_{F}^{1/2}} = \frac{u^\star}{1-0.62 u^\star }.
 \label{key0}
\end{equation}
In the regime $L>L_{c}= \sqrt[3]{\frac{2\pi}{n}}$, the superconductivity is suppressed, i.e. $T_c=0$, according to the solution to the following equation:
\begin{align}
& 2\,D\,\frac{N^{bulk}(\epsilon_F)}{(\epsilon_{F}^{bulk})^{-1/2}}\left(1+\frac{2}{3}\frac{\pi}{n L_{cr}^{3}}\right)^{1/3}\left(1+\frac{2}{3}\frac{\pi}{n L_{cr}^{3}}\right)^{-1/3} \nonumber\\
&~~~~~~~=2\,D\,N^{bulk}(\epsilon_F)(\epsilon_{F}^{bulk})^{-1/2}\nonumber\\
&~~~~~~~=\frac{u^\star}{1-0.62 u^\star }.
\label{key}
\end{align}
This therefore shows that there exists no value of film thickness at which superconductivity is suppressed, for systems where the Thomas-Fermi screening is large compared to the wave vector of the mediating phonon, in the absence of an external electric field. The ultimate reason for this effect is the cancellation, in $\lambda$, between the confinement-induced modification of the density of states at the Fermi level and the confinement-induced modification of the Fermi energy. The latter is contributed by the Bardeen-Pines electron-phonon matrix element that includes screening.


For completeness, we shall also consider the regime $L<L_{c}= \sqrt[3]{\frac{2\pi}{n}}$.
Using Eq. \eqref{Fermi_L_large} and Eq. \eqref{Fermi_L_small} inside Eq. \eqref{key0}, we obtain:
\begin{equation}
4 D \frac{V m}{\sqrt{2}\pi^{3/2}}\frac{\sqrt{L_{cr}n}}{\hbar^{2}} \left[\frac{\hbar^{2}}{m}\left(\frac{(2\pi)^{3}n}{L_{cr}}\right)^{1/2}\right]^{-1/2}=\frac{u^\star}{1-0.62 u^\star }.
\end{equation}
Hence, in this case there exists a critical film thickness $L_{cr}$ at which superconductivity is suppressed. Solving for $L_{cr}$ we obtain the following experimentally testable expression:
\begin{equation}
L_{cr} = \frac{2^{-3}\pi^{9}}{D^{4}}\frac{n^{-1}}{V^{4} m^{6}}\left(\frac{u^\star}{1-0.62 u^\star }\right)^{4}
\end{equation}
which, however, may be difficult to confirm experimentally since it is valid for $L<4-5${\AA} which is already at the 2D limit.

Using, instead, the relation for $\lambda$ valid in the regime where $k_{TF} \ll q$, Eq. \eqref{lambda no screening}, we obtain:
\begin{equation}
2\,D'\,N^{bulk}(\epsilon_F)\left(1+\frac{2}{3}\frac{\pi}{n L_{cr}^{3}}\right)^{1/3}=\frac{u^\star}{1-0.62 u^\star }.
\label{key2}
\end{equation}
from which the critical thickness at which superconductivity is suppressed follows as:
\begin{equation}
L_{cr} = \sqrt[3]{\frac{2 \pi}{3} \left[\left( \frac{1}{2\,D'\,N^{bulk}(\epsilon_F)}\frac{u^\star}{1-0.62 u^\star} \right) -1\right]^{-1}}.
\label{key}
\end{equation}
It should be noted that the main reason here why $L_{cr}$ is not zero is that the Coulomb repulsion parameter $u^\star$ is not zero. This can be confronted with the result obtained in the weak-coupling BCS limit, where $u^\star=0$ and, therefore, $L_{cr}=0$ in agreement with Ref. \cite{Travaglino}.

\subsection{Thickness-dependent effect of Coulomb repulsion}
In the above, we developed a microscopic theory of superconductivity in thin films that also accounts for screening, however, we neglected the dependence of the Coulomb parameter $u^\star$ on the film thickness $L$. Within the Tolmachev-Morel-Anderson pseudopotential theory \cite{Morel,Tolmachev}, the Coulomb parameter is given by
\begin{equation}
u^\star=\frac{N(\epsilon_F)U}{1 + N(\epsilon_F)U \ln (\epsilon_F/\omega_c)} \label{Coulomb}
\end{equation}
where $U$ is is a double average of the
direct Coulomb repulsion on the Fermi surface \cite{Carbotte2003}. For the most materials of experimental and technological relevance, the range of $u^\star$ goes from 0 to $0.2$. In this range, clearly, $u^\star$ is a monotonically increasing function of $N(\epsilon_F)$. In turn, according to Eq. \eqref{Fermi_L_large0}, $N(\epsilon_F)$ is a monotonically decreasing function of the film thickness $L$. Since we can safely neglect the much weaker $L$-dependence of the logarithmic factor $\ln (\epsilon_F/\omega_c)$, we conclude that $u^\star$ is a monotonically decreasing function of the film thickness $L$. In other words, upon decreasing the film thickness $L$, the quantum confinement causes the Coulomb repulsion parameter $u^\star$ to increase, which is clearly a detrimental effect for superconductivity. 

\subsection{Thickness-dependent critical electric field for superconductivity suppression}
Within the microscopic theory, the superconductivity will be suppressed when a critical EF is applied which is large enough to break the Cooper pairs, i.e. \cite{Patino2021,Ashcroft}
\begin{equation}
E_{cr}=\frac{2 \Delta}{e\,\xi}, 
\end{equation}
where $\Delta$ is the BCS energy gap, which also represents the binding energy of a Cooper pair, $e$ is the electron charge, and $\xi$ is the coherence length, which is a measure of the spatial distance between two electrons forming a Cooper pair.
Within a simplified BCS picture where screening is not taken into account, in the regime $L>L_c$ we can write:
\begin{equation}
\Delta = \frac{2\hbar \omega_{D}}{\exp[\frac{1}{N(\epsilon_{F})U}]-1}.
\end{equation}
Upon substituting Eq. \eqref{Fermi_L_small0} we obtain
\begin{equation}
E_{cr} = 4\frac{\hbar \omega_{D} }{e\, \xi}\left[\exp\left(\frac{1}{N^{bulk}(\epsilon_{F}) \left(1 + \frac{2}{3} \frac{\pi}{n L^3}\right)^{1/3} U}\right)-1\right]^{-1}.
\end{equation}
From this equation, we see that $E_{cr}$ increases as $L$ decreases, hence the Cooper pair binding energy increases as $L$ decreases and a higher value of $E_{cr}$ is needed to suppress the superconductivity.

In the regime $L<L_c$, instead, we need to use Eq. \eqref{Fermi_L_large}, leading to 
\begin{equation}
E_{cr} = 4\frac{\hbar \omega_{D} }{e\, \xi}\left[\exp\left(\frac{1}{ 2\frac{V m \sqrt{L n}}{\sqrt{2} \pi^{3/2}\hbar^2}U}\right)-1\right]^{-1}.
\end{equation}
In this regime, instead, we see that, upon further decreasing $L$, the electric field needed to break the Cooper pairs, $E_{cr}$, increases.
This is consistent with the fact that the $T_c$ in the regime $L<L_c$ decreases with further decreasing the film thickness $L$ \cite{Travaglino}.
However, this is a simplified way of evaluating the critical field, which does not consider the effect of the Thomas-Fermi screening, and does not take into account the fact that the Coulomb repulsion parameter $u^\star$ also changes with confinement.

The more comprehensive approach is to evaluate the critical field in terms of the Cooper pair binding energy and then in terms of $T_c$ as:
\begin{equation}
E_{cr}=2 \frac{\Delta}{e\, \xi} \approx 2 \cdot \frac{1.764}{e\, \xi} \, k_{B}T_{c} \label{field gap}
\end{equation}
and use the Allen-Dynes $T_{c}$ formula Eq.\eqref{allenformula} evaluated from Eliashberg theory together with Eq.\eqref{lambda} for $\lambda$. 
In particular, for $L>L_c$ and in the case of negligible Thomas-Fermi screening, we have the following dependence of the electron-phonon coupling upon the film thickness $L$:
\begin{equation}
\lambda \sim  \left(1+\frac{2}{3}\frac{\pi}{n L^{3}}\right)^{1/3}.
\end{equation}
Upon substituting inside the Allen-Dynes formula, we thus get:
\begin{align}
E_{cr} &= \frac{C}{\xi} \exp \left[ - \frac{1.04\left[1+\lambda\left(L\right)\right]}{\lambda\left(L\right)-u^\star(L)\left[1+0.62 \lambda\left(L\right)\right]}\right],\nonumber \\
\lambda(L)&=2D N^{bulk}(\epsilon_{F})\left( 1+\frac{2}{3}\frac{\pi}{n L^{3}}\right)^{1/3}
\label{crucial}
\end{align}
where $C =  2.94 k_B \,f_1\,f_2\,\omega_{log}$. Here, $\lambda(L)$ decreases monotonically as $L$ decreases (see above), and $u^\star(L)$ also decreases monotonically with $L$ with basically the same law as $\lambda$ because $u^\star \propto N(\epsilon_F)$. Since $u^\star < \lambda$, the overall trend of $E_{cr}$ with $L$ is still dominated by $\lambda(L)$ inside the exponential. Hence, since the exponential factor in Eq. \eqref{crucial} is well-known to be a monotonically increasing function of $\lambda$, cfr. Ref. \cite{PhysRevB.12.905}, for all known materials, the above equation implies that a lower critical value of EF to suppress superconductivity is required upon decreasing the film thickness, which is at odds with experimental observations \cite{Giazotto2018}.

In the case where screening is important, $k_{TF} \gg q$, we have shown that there is an exact cancellation of effects such that $\lambda$ is independent of $L$ due to the competing dependencies on $L$ brought about by $N(\epsilon_F)$ and by the Thomas-Fermi screening, respectively. Hence, in this case Eq. \eqref{crucial} becomes:
\begin{align}
E_{cr} &= \frac{C}{\xi} \exp \left[ - \frac{1.04\left[1+\lambda\right]}{\lambda-u^\star(L)\left[1+0.62 \lambda\right]}\right],\nonumber \\
\lambda&=2D N^{bulk}(\epsilon_{F})\left( \epsilon^{bulk}_{F}\right)^{-1/2}
\label{crucial final}
\end{align}
that is, the dominant dependence on the film thickness $L$ is the one contributed by the Coulomb repulsion parameter $u^\star(L)$ inside the exponential factor, via Eq. \eqref{Coulomb} and Eq. \eqref{Fermi_L_large0}. Since $\Delta \sim E_{cr}$ is a monotonically decreasing function of $u^\star(L)$, and $u^\star(L)$ is a monotonically decreasing function of $L$, this result directly implies that  $E_{cr}$, i.e. the critical EF value to suppress superconductivity, decreases upon decreasing the film thickness $L$. This is the most important finding of the present paper, which explains some experimental observations of Ref. \cite{Giazotto2018}.
While in the above discussion we neglected the possible dependence of the coherence length $\xi$ on $L$, which cannot be predicted a priori since it is a function of the actual film fabrication process, we expect this dependence to be sub-dominant with respect to the $L$-dependence of the exponential factor in Eq. \eqref{crucial final}.

This procedure allows for the exact calculation of the dependence of $E_{cr}$ upon $L$, which is left for future work using input from ab-initio methods \cite{EPW}.

Material and temperature dependence of the critical field
To summarize, the main result of our theory is Eq. \eqref{crucial final}
with the thickness-dependent Coulomb parameter given by Eq. \eqref{Coulomb} evaluated with Eq. \eqref{Fermi_L_large0}.

The dependence of the Coulomb parameter $u^\star$ on the film thickness $L$ controls the thickness-dependence of the critical field $E_{cr}$. This is a \emph{universal} dependence on $L$, valid for all materials and independent of composition.


In the above equation for the critical EF, a possibly material-dependent parameter is the coherence length $\xi$. The latter is given by \cite{deGennes}:
\begin{equation}
    \frac{1}{\xi}=\frac{1}{\xi_0}+\frac{1}{\ell}
\end{equation}
where $\xi_0$ is the intrinsic (Pippard) coherence length, and $\ell$ is the mean free path. Thin films, such as those used in the supercurrent field effect devices, have a microstructure characterized by microcrystallites, the size of which sets the value of $\ell$. Since, typically, $\ell \ll \xi_0$ (because $\xi_0$ can be tens or hundreds of nanometers), the coherence length $\xi$ is controlled by $\ell$, and, hence, by the fabrication process.

The experimental observation that $E_{cr}$ is within two orders of magnitude for different materials, as shown in Fig. 3 of Ref. \cite{ruf}, is explained by the fact that $E_{cr} \sim \Delta(L)/\xi$, where $\Delta(L)$ is the thickness-dependent energy gap. Because the thickness dependence of $\Delta(L)$ is the same for all materials, the only material dependence is that of the energy gap of the bulk material. Since (i) $\Delta \simeq 1.76 k_B T_c$ for the bulk material, where $T_c$ varies for conventional superconductors within a factor of 20 \cite{Bussmann-Holder}, and (ii) the coherence length $\xi$ in thin films, which is set by $\ell$, is estimated to vary by at most an order of magnitude (1-10 nm), depending on the fabrication process, it is clear that $E_{cr}$ can vary within two orders of magnitude.

We also notice that, as a matter of fact, the above thickness dependence of the critical electric field $E_{cr}$ is the same at all temperatures. This is evident from the above Eq. \eqref{crucial final}, and from the fact that our confinement model for the quasiparticles is entirely temperature-independent (since it is a purely quantum effect).

Finally, since the coherence length $\xi$ is a measure of the physical separation between two electrons forming a Cooper pair, it is clear that the above theory is applicable only to systems where $\xi < L$.

\section{Conclusions}
In summary, we have developed a microscopic theory of phonon-mediated superconductivity in thin metallic films based on a recent quantum confinement model \cite{Travaglino,Travaglino_2022}, that also accounts for the Thomas-Fermi screening effects and the Coulomb repulsion. 
The theory is able to predict the experimental observation \cite{Giazotto2018} that a lower critical electric field is needed to suppress superconductivity as the thickness $L$ of the film is decreased.

The theory presents a possible basis to qualitatively explain and understand the field-effect on supercurrent in thin films. In particular, to explain the suppression of superconductivity as a function of the film thickness and the relationship between the critical electric field and the film thickness. A crucial role in the ability of the theory to explain the thickness-dependent suppression of superconductivity in thin films is played by the synergy between the Thomas-Fermi screening in the electron-phonon matrix element, and electron-electron Coulomb repulsion. In particular, thanks to the Thomas-Fermi screening and its dependence on the thickness-dependent Fermi energy, the electron-phonon coupling $\lambda$ becomes independent of film thickness due to an exact cancellation of the opposite effects. This leaves, as the only active thickness dependence, the one of the Coulomb parameter $u^\star$, which increases upon decreasing the film thickness, thus lowering the energy gap as the thickness decreases.
Furthermore, we have estimated the critical value of the external electric field, $E_{cr}$, required to suppress the superconductivity under these conditions. When both the Thomas-Fermi screening and the Coulomb repulsion, with their respective dependencies on the film thickness, are duly taken into account, $E_{cr}$ decreases upon decreasing the thickness of the thin film, in agreement with the known phenomenology \cite{Giazotto2018,Orus2021}.

All in all, these results represent a step forward for the rational and quantitative optimization of the field-effect transistors and for electric-field gated
quantum materials, and it can be extended in a future work to ultrathin films of few atomic layers \cite{Chiang}. 
Furthermore, a wealth of new fundamental laws relating physical quantities have been presented, which are testable in future experiments.

\section*{Acknowledgments} 

A.Z. gratefully acknowledges funding from the European Union through Horizon Europe ERC Grant number: 101043968 ``Multimech'', from US Army Research Office through contract nr. W911NF-22-2-0256, and from the Nieders{\"a}chsische Akademie der Wissenschaften zu G{\"o}ttingen in the frame of the Gauss Professorship program. V. M. F. gratefully acknowledges support by the European Cooperation in Science and Technology via COST Action CA21144 (SUPERQUMAP) and fruitful discussions with F. Giazotto, R. Córdoba, J. M. De Teresa, A. Di Bernardo, M. Cuoco, and G. Ummarino, L. Martinoia, A. Amoretti.

\bibliography{refs}

\end{document}